\documentclass[11pt,twoside]{article}

\usepackage{asp2006}
\usepackage{epsf}
\usepackage{lscape}
\usepackage{graphicx}

\begin{document}
\title{ Reconstruction and Analysis of Component Spectra of Binary and Multiple Stars} 
\author{K. Pavlovski$^1$ and H. Hensberge$^2$}  
\affil{$^1$ Department of Physics, University of Zagreb, Zagreb, Croatia \\
       $^2$ Royal Observatory of Belgium, Ringlaan 3, Brussels, Belgium} 

\begin{abstract}
In the last two decades about a dozen methods were invented which derive, 
from a series of composite spectra over the orbit, the spectra of individual 
components in binary and multiple systems. Reconstructed spectra can then be 
analyzed with the tools developed for single stars. Eventually this has created 
the opportunity for chemical composition studies in previously inaccessible 
components of binary stars, and to follow their chemical evolution, an important 
aspect in understanding evolution of stellar systems. First, we review new 
developments in techniques to separate and reconstruct individual spectra, 
and thereafter concentrate on some applications. In particular, we emphasize 
the elemental abundance studies for high-mass stars, and present our recent 
results in probing theoretical evolution models which include effects 
of rotationally induced mixing.
\end{abstract}

\section{Introduction}  
Shifts of the spectral lines due to the orbital motion of components
in binary and multiple systems are essential for determination of
stellar masses.
Due to the intermingling of several diluted components, spectra of
binary and multiple stars are often complex.
Typically lines overlap in the course of the orbital
cycle, which makes measurements of the radial velocities difficult
and less accurate than desired for the determination of reliable
stellar masses (Andersen 1991, Torres, Andersen \& Gim\'{e}nez 2009),
and prevents a straightforward astrophysical analysis of the
composite spectra.
  However, in spite of all
difficulties encountered progress has been made, in  particular for
Algol type systems (c.f.~Tomkin 1989). 

Tremendeous advance in  astronomical instrumentation and in
efficiency and linearity of detectors  have
 prompted the development of new analysis techniques. 
These are now able to exploit all the intermingled Doppler shift
information present in the spectra.
New methods are offering incomparable opportunities, 
both for deriving precise fundamental stellar quantities
 and for investigating 
 the chemical evolution of binary and multipe stars.

\section{Reconstruction techniques}

In the last two decades about a dozen methods were invented which derive,
from a series of composite spectra over the orbit, the spectra of individual
components in binary and multiple systems.
Herewith, these reconstruction methods are divided into 
  three basic groups: (i) spectral separation, 
(ii) spectral disentangling, and (iii) spectroastrometric splitting:

\begin{itemize}
\item Spectral separation: The spectra of components are reconstructed from
a time series of composite spectra for corresponding radial velocities and 
light factors of the components. 

\item Spectral disentangling: Self-consistent solution of a time series of
composite spectra which gives individual spectra of components and
set of orbital elements. No {\em a priori} knowledge of RVs is needed.

\item Spectroastrometric splitting: Technique which utilises the spatial
information present in a longslit spectrum. It can not only be used
to detect binary systems, it can also deconvolve the observed
spectrum into the individual spectra of its components.
\end{itemize}

\subsection{Spectral separation}

{\em Doppler tomography} was the first successful reconstruction of individual spectra
of the components from a time series of composite spectra that did not rely on
template spectra (Bagnuolo \& Gies 1991). The reconstruction is made through an 
iterative least-squares solution. The radial velocities of the components 
have to be known and are used as input data. Recently, the 11th paper in the series
on tomographic separation was published by Penny, Ouzts \& Gies (2008). 
Further informations on this and other methods,  were given in review papers by
Gies (2004), and Hensberge \& Pavlovski (2007).

{\em Subtraction procedure} has shown to yield resonable results for systems
containing a cool giant and a hotter main-sequence star (Griffin \& Griffin 1986).
 Recently, the authors have published their 15th paper in the series on composite spectra
(Griffin \& Griffin 2009).

{\em Doppler differencing} introduced by Ferluga et al.\ (1997) is the most
straightforward method of direct subtracting, in which two spectra of a binary
obtained in the phases of opposite extremes of RVs are used. Recently, Lee et al.\
(2008) revived this method in a successful separation of an extremely interesting
quadruple system containg two eclipsing pairs.

\subsection{Spectral disentangling}

{\em Spectral disentangling in wavelength domain} or {\em disentangling in velocity-space} 
was invented by Simon \& 
Sturm (1994).
It is based on solving the 
matrix equation $A  x = y$, where in vector $y$ all
observed spectra are stored, and $x$ contains the spectra of the components. Matrix
$A$ has elements (blocks) corresponding to the Doppler shifts and light factors.
The block $A_{jk}$  for each observed spectrum, where $k$ identifies 
the component and $j$ the time of observation, transfers $x_k$ into the $j$-th
subvector of $y$.

The structure of blocks $A_{jk}$ is rather simple.
These are diagonal matrices with the diagonal shifted according to Doppler shift 
of component $k$ at time $t_j$ and multiplied by a light factor.
Their size is determined by the number of bins in the spectral
interval and the maximum size of the Doppler shifts. Therefore, vectors $x_k$ must
be somewhat longer than the subvectors $y_j$.
Solving this
matrix equation for $x$ requires the inverse of $A$. Since matrix $A$ is not a square
matrix, and since it may be rank deficient,  a singular value
decomposition must be used. The system is an over-determined system of linear equations 
(more equations than unknowns) when the input consists of more spectra than stellar
components, 
and {\em least squares solution} is required such that 
it minimizes the norm of the residuals $r = || A  x - y ||$. Simon \& Sturm (1994) were the first
who beside reconstruction of individual spectra also optimized a set of orbital
elements of a binary system. In this sense {\em spectral disentangling} was formulated
and is distinguished from {\em spectral separation} techniques (Sect.\ 2.1)

Wavelenght-domain spectral disentangling has several interesting features as emphasized
by Iliji\'{c} (2004), (i) original sampling of observed spectra can be preserved since
the input spectra must not be resampled onto same wavelength grid, (ii) weight can be
assigned to each pixel in the observed spectrum, (iii) in consequence of (ii) some parts
of observed spectra can be omitted, like interstellar lines and bands, telluric lines,
emission lines,  blemishes and other technical imperfections which
are not cleaned in the reduction of the spectra. 
An important drawback of the approach in (logarithmic) wavelength-domain, or velocity
domain, is the prohibitively large CPU time and memory storage involved, due to
the coupling of the huge number of equations ($N_{\rm obs} \times N_{\rm pix}$).
Therefore,  
this method should be used only in the gridding mode or in the separation mode with
known orbital elements.

{\em Spectral disentangling in Fourier domain} or {\em Fourier disentangling} 
was developed by Hadrava (1995) and resolves elegantly the CPU time overhead of the
wavelength-domain approach. 
The equations for the discrete
Fourier transforms of the spectra reduce to a huge number 
 ($(N_{\rm pix}/2) + 1$) of small sets of $(N_{\rm obs})$ complex
equations with only few $(k)$ unknowns.
  When dealing with long spectral windows
like in the case of \'{e}chelle spectra this issue is of essential importance.
Weights can be given to specific Fourier components (e.g.~eliminating the influence
of badly constrained low-order components) and to complete spectra, but -- nothing
comes for free -- weight can  not be given to pixels.
With the cyclic character of finite Fourier transforms, there should be ideally
sufficient continuum at the same level at both ends to avoid `end-of-range' effects
(Iliji\'{c} et al.\ 2001a,b) from spectral features leaving and entering the data 
set due to Doppler shifting.
 An optimization is also performed for the orbital
elements, and measurements of RVs are bypassed, as in the method of Simon \&
Sturm (1994). Detail explanations and mathematical expositions can be found
in Hadrava (2004, 2009).

{\em Iterative Doppler differencing} is a method developed by Gonz\'{a}lez \& Levato (2006)
on some premises used earlier by Marchenko, Moffat \& Eenens (1998). This is an iterative scheme,
using alternatively the spectrum of one component to predict the spectrum of the other one. 
In each step, 
the contribution of one component to the observed spectra is eliminated, RVs are measured
for the other component, and applying the related Doppler shifts a new estimate of the
spectrum of the other component is obtained.

{\em Non-linear least-squares fitting} was applied by Harries et al.\ (2003), and indipendently
by Napiwotski et al.\ (2003). The idea is to fit composite spectra for two components for a given
set of orbital parameters or RVs. Harries et al. {2003) have used a genetic algorithm for
minimization of the residuals between the observed and model composite spectra, while
Napiwotzki et al.\ (2003) in their code {\sc fitsb2} are using Simplex (c.f.~Drechsel \&
Nesslinger, this Volume, for the recent use of this code).

\subsection{Spectroastrometric splitting} 

{\em Spectroastrometic splitting}  method is based on the positional measurements,
hence its name. Conceptually, the centroid of a flux distribution to a
fraction of pixel is measured on a long slit spectrum. 
The distribution of the light at a specific wavelength in the extraction slit changes
locally if one component contributes to a spectral feature stronger than the
other component, relative to nearby continuum.
 The method was
developed by Bailey (1998a), Garcia et al.\ (1999), Takami et al.\ (2001), and Porter,
Oudmaijer \& Baines (2004). It proved to be a powerful technique providing information on the flux
distribution at milliarcsecond scales. In a  recent application of the
method of {\em spectroastrometric splitting} the spectrum of $\beta$ Cep was split
into its constituent spectra by Wheelwright, Oudmaijer \& Schnerr (2009), which enabled
them to disclose that the fainter secondary exhibits signatures of a classic Be star.

\section{Spectral disentangling in practice}

Even in its first applications for deriving the orbital elements, spectral
disentangling ({\sc spd}) techniques 
were found to be superior to other methods (Sturm \& Simon 1994, Simon et al.\ 1995, Harmanec
et al.\ 1997). Recently, Southworth \& Clausen (2008) have examined this issue in
 detail. They measured RVs  by 
fitting  double Gaussians, and by calculating 1D and 2D cross-correlation functions ({\sc ccf}).
The  derived orbital elements
are compared to those derived by {\sc spd} in wavelength-domain. It was found that
the line blending is a more severe obstacle for cross-corelation. 2D cross-correlation did not
improve results, probably due to the spectral similarity of the component stars.  
 It is encouraging that the orbits measured from {\sc spd} were the
most internally consistent, and did not require any corrections for line blending. 
However, Southworth \& Clausen have found that {\sc spd} suffers from the presence
of many local maxima (minima) in the parameter space, as is common in non-linear
multi-parameter optimization.

\begin{figure}[] \centering
\begin{tabular}{cc}
\includegraphics[width=55mm]{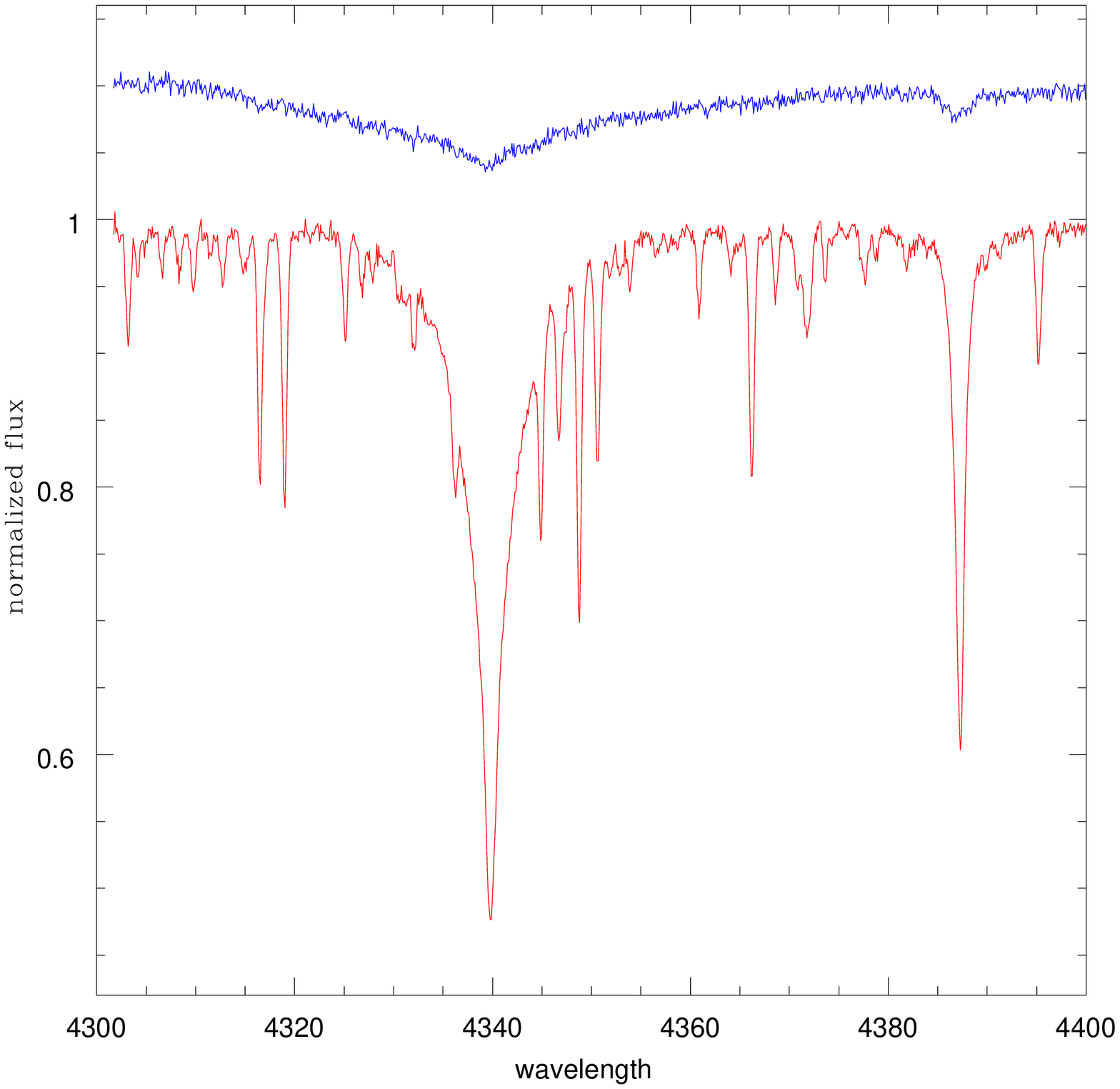} & 
\includegraphics[width=55mm]{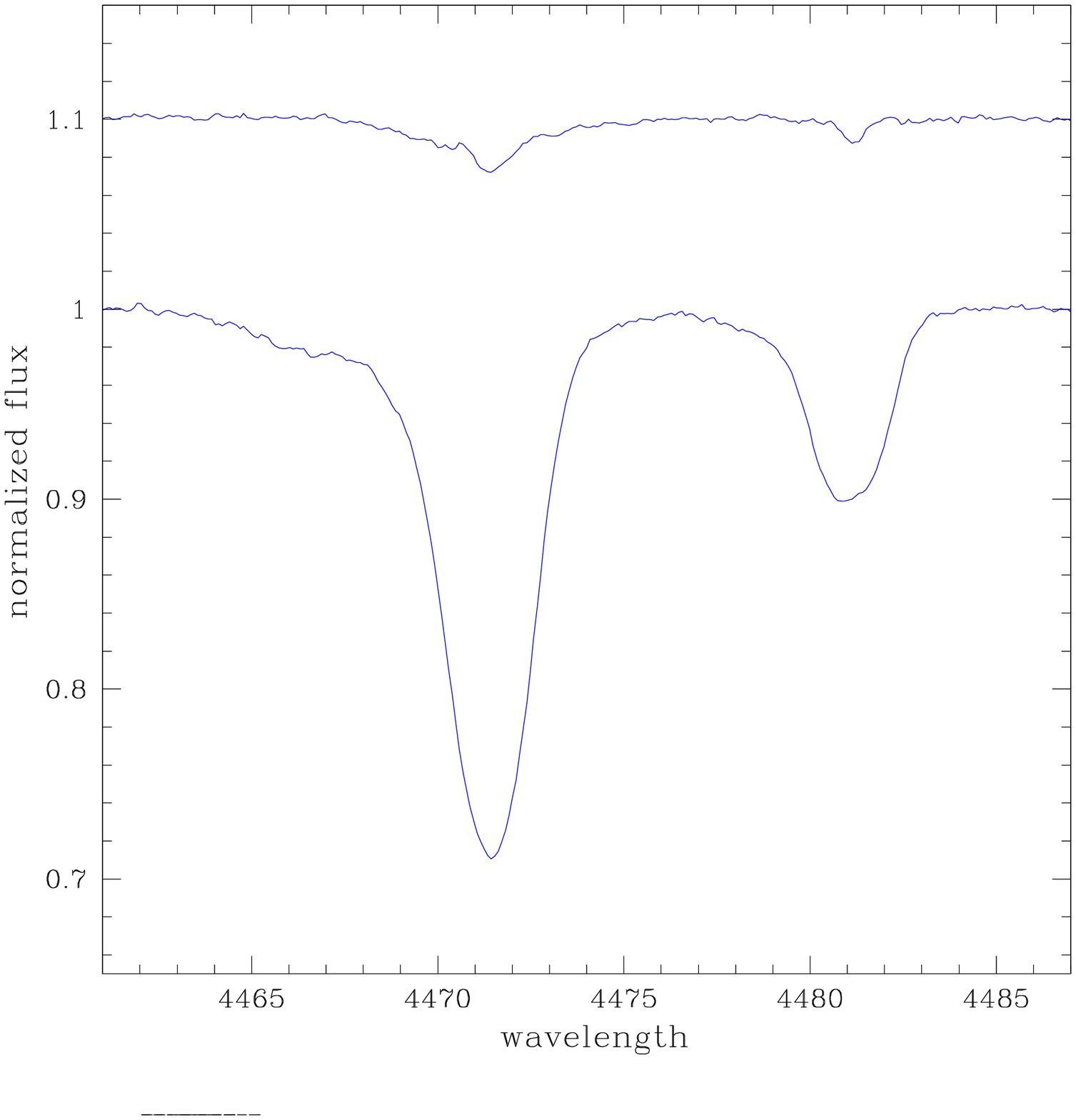} \\
\end{tabular}
\caption{Detection of the faint companions in binary systems by Fourier disentangling, 
V621 Per in the open cluster h Per (left pannel) turns to be SB2 system (Southworth et al.\ 2009,
in preparation), and V380 Cyg (right pannel, after Pavlovski et al.\ 2009). Spectra of the secondary
components are shifted arbitrary for clarity.}
\end{figure}

Although the use of the disentangling gets well-spread in the astronomical community,
often RVs are measured afterwards using the component spectra as templates, for use in
software combining light curve or astrometric information with spectroscopic information.
One should be aware of the fact that calculating the orbital elements from RVs measured in this
way will differ somewhat from the ones determined simultaneously with the  component spectra.
This  was shown by Iliji\'{c} (2004) who performed numerical experiments
with time series of
synthetic composite spectra  calculated from the two input spectra, and
given RVs defined by  orbital elements. 
{\sc ccf} was performed both with
 input and disentangled spectra of the components
as templates. In no case did {\sc ccf} return the input pair of RVs.
This is due to the limited precison of the orbital elements in the
presence of noise in the input spectra, and the noise progresion into the
disentangled component spectra which {\em correlates} with the noise in the
orbital elements. Therefore, the combination of spectroscopic information with other
types of data should no longer be based on RVs but directly on the norm of the
residuals in the disentangled algorithms. Existing codes should be adapted in this 
way to ascertain optimal results.
 Some of these issues were investigated using synthetic data by Hynes \& Maxted
(1998). This problem deserves further and thorough study.

Undulations or spurious patterns are a long-standing issue. 
In a comprehensive study, Hensberge et al.\ (2008) have found a way to
 remove or reduce these spurious undulations in a self-consistent way.
Also, hints are given to reduce
quasi-degeneracies by obtaining better-conditioned data sets.
Undulations are understood to be due to 
indeterminacies in the set of equations (low-frequency Fourier components) in the
absence of sufficiently strong time-dependent dilution of spectral lines and/or
to imperfect normalisation of the input spectra. 

{\sc spd} has been shown to be capable to  detect  faint companions. In AR\,Cas
the secondary component, which is about 4\,mag fainter than the primary, 
was detected by  Holmgren  et al.\ (1999). 
Spectroscopic diagnostics were possible on disentangled secondary
component of V380\,Cyg, which contributes only about 6\% of the total light of
binary system (Pavlovski et al.\ 2009). With the application of {\sc spd} Southworth
et al.\ (2009, in prep.) have turned the SB1 system V621\,Per into an SB2 system. 
The light ratio is  22 and the detection of the faint component is made
in the strong and broad hydrogen lines of the late-B secondary, explaining
the failure of  Southworth et al.\ (2004b) to detect it by  {\sc ccf} (Fig.~1).
The spectrum of the hot but faint subdwarf in FY\,CMa was isolated by tomographic
spearation in Peters et al.\ (2008).
In a detailed study of the prototypical mass-transferring system Algol, Fourier
disentangling of more than 80 high-resolution
and high S/N spectra (obtained using {\sc fies} at the Nordic Optical Telescope)
has revealed all three components, of which the K1 giant secondary star
contributes less than 2\% (Pavlovski, Kolbas \& Southworth, this Volume).
This is the faintest star ever detected by {\sc spd}.

Disentangling was found to be very efficient in detecting
new binary stars; for example the survey of the open cluster NGC\,2244 by Mahy et al.\ (2009), X-Mega
targets in the Carina nebula by Rauw et al.\ (2009),
and the spectroastrometric survey
of PMS stars by Bailey (1998b).

{\sc spd} returns the orbital elements of the binary systems with a higher accuracy than
achieved by other methods (Holmgren et al.\ 1999; Hensberge et al.\ 2000;
Southworth \& Clausen 2007; Baki\c{s} et al.\ 2007; Baki\c{s} et al.\ 2008; Mayer et al.\ 2008;
Pavlovski \& Southworth 2009; Pavlovski et al.\ 2009). Disentangling
techniques were succesfully applied in the studies of triple systems (Fr\'{e}mat et al.\ 2005;
Baki\c{s} et al.\ 2008; Hareter et al.\ 2008), and quadruple systems where different
methods were applied (Gonz\'{a}les et al.\ 2006;
Lee et al.\ 2008; Harmanec et al. 2007).
Algol (semidetached) systems were also studied; RY\,Per by Boyajian et al.\ (2004),
$\delta$\,Lib by Baki\c{s} et al.\ (2006),
u\,Her by Hilditch (2007), RY\,Sct by Grundstrom et al.\ (2007), and Algol itself by
Pavlovski, Kolbas \& Southworth (this Volume). First applications to chromospherically-active
binaries were made by Diaz et al.\ (2008), Wang et al.\ (2009), Do\u{g}ru et al.\ (2009),
and to W\,UMa-type systems by  \"{O}zkarde\c{s} et al.\ (2009).

 Fourier disentangling made possible  the separation of a focussed
wind component of the black hole binary X-ray source Cyg\,X-1 (Yan, Lin \& Hadrava  2008).
Linder et al.\ (2008) applied Doppler tomography to produce a Doppler map of Plaskett's
star (HD\,47129) showing the wind interaction in the system.

{\sc spd} is an especially attractive tool for distance
measurements, since accurate effective  temperatures can be
derived from disentangled spectra.
An improved accuracy in the distance to the Rosette Nebula cluster NGC\,2244 was achieved
in the study of the eclipsing and SB2 system V578\,Mon (Hensberge et al.\ 2000). In similar studies
distances to the open clusters were provided for the Pleiades (Zwahlen et al.\ 2004;
Groenewegen et al.\ 2007),
NGC\,188  (Meibom et al.\ 2008) and NGC\,6791 (Grundahl et al.\ 2009).
{\sc spd} of  binaries in Local Group galaxies also helps in the proper scaling of
cosmological distance ladder (Ribas et al.\ 2005; Bonanos 2009; North, Gauderon \& Royer 2009).

\begin{figure}[] \centering
\begin{tabular}{cc}
\includegraphics[width=58mm]{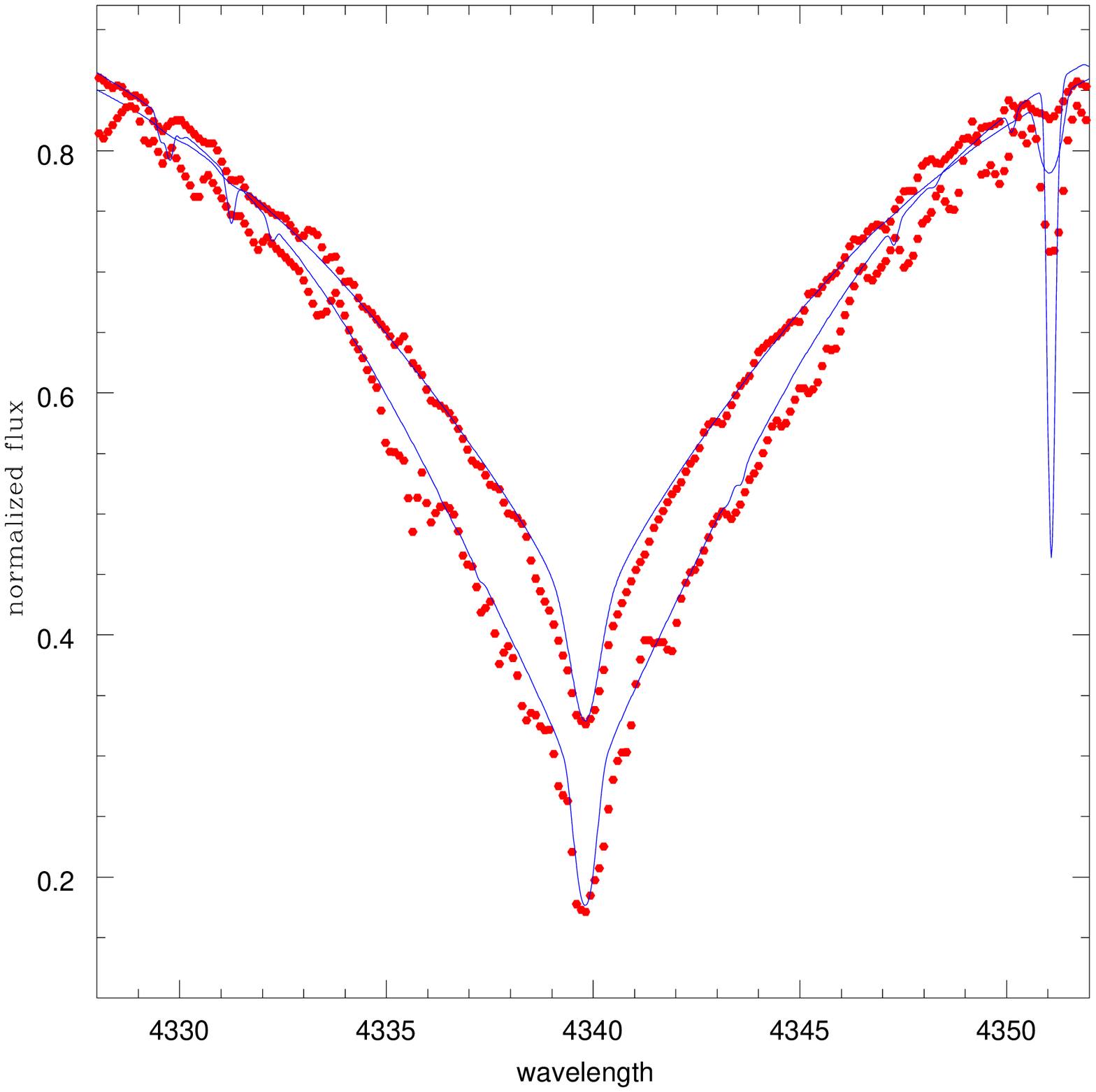} &
\includegraphics[width=58mm]{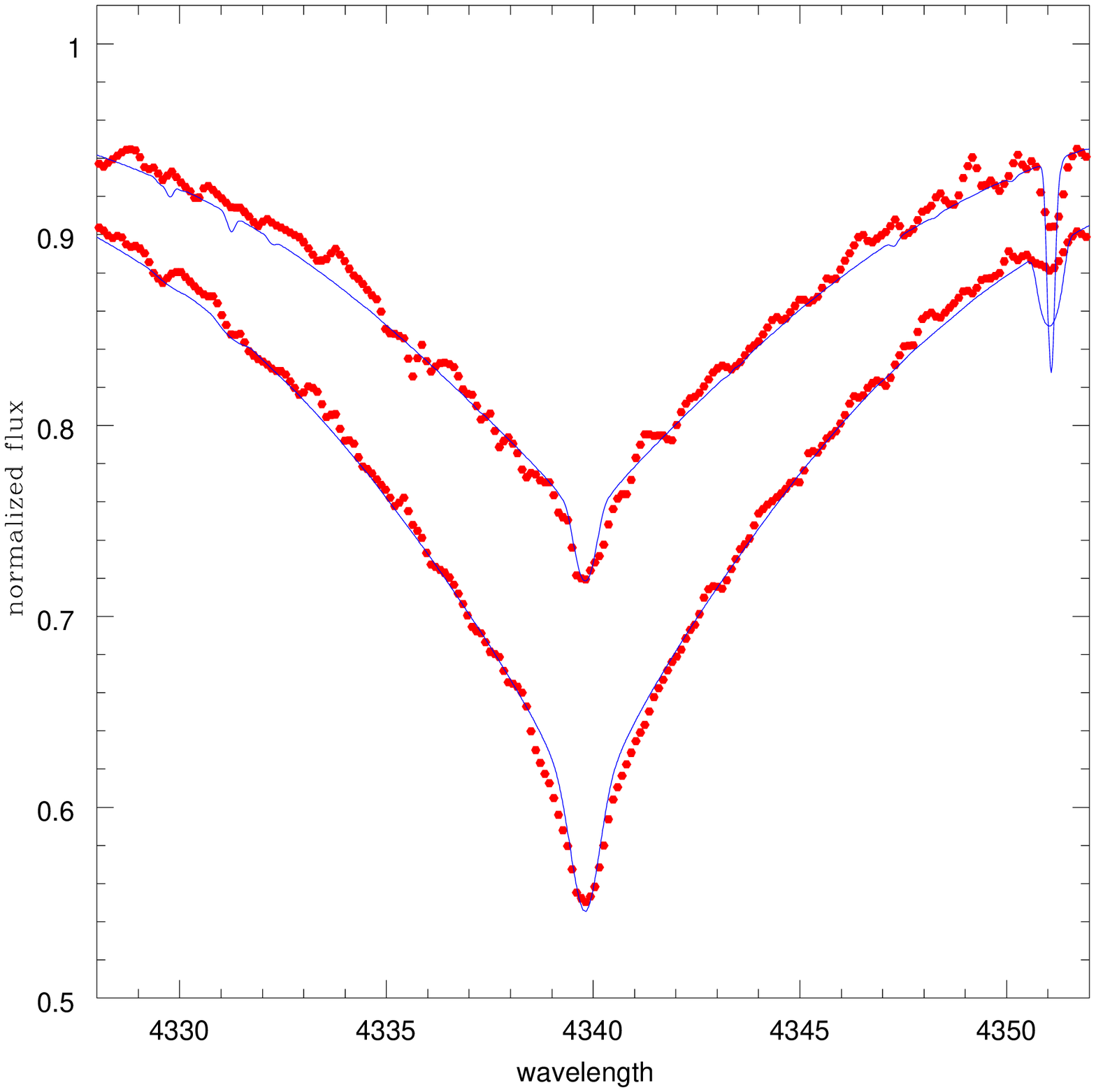} \\
\end{tabular}
\caption{Deriving effective temperatures for the components of V615\,Per. Spectra around H$\gamma$
are renormalised with light factors from the light curve analysis (left panel). Both effective
temperatures and light factors are derived by fitting separated (disentangled) profiles (right
panel).}
\end{figure}

\section{Renormalisation of disentangled component spectra}

Separate individual spectra of the components of multiple systems are either in the common continuum
of the system, or in an arbitrary (generic) mode. If the light factors
change significantly between spectra (particularly if eclipse spectra have been obtained)
then the separation will directly give individual spectra with correct continuum levels.
This requires that the light factors must either be determined from the time-dependent
dilution of the spectral lines (`line photometry', Hadrava 1997), or must be estimated
from external sources, usually light curves (but see below).

Renormalisation is needed if spectroscopic diagnostics and abundances are to
be derived from disentangled spectra. Depending on the nature of the binary/multiple
system under consideration, several approaches are possible:

\begin{itemize}
\item For eclipsing binaries, light factors may be available from time-independent
 dilution of spectral lines or from (eclipse) light curves, leading straightforwardly
 to undiluted disentangled spectra (e.g.~Hensberge et al.\ 2000).

\item If the system is not eclipsing, some {\em physical considerations} still can be used
 to renormalise individual disentangled spectra. In the study of the non-eclipsing
 triple system DG\,Leo, Fr\'{e}mat et al.\ (2005) successfully used the very deep  Ca\,{\sc ii}
 K-line. The requirement that the core should not cross the zero-point of any of the
 component spectra (all of similar spectral type) imposed very strong constraints
 on the light factors.

\item Separated or generic disentangled spectra contain information on the intrinsic
 spectra. Just as in the way in which information on effective temperatures and surface
 gravities are extracted from renormalised spectra, it is possible to recover also
 light factors by constrained multi-parameter line-profile fitting of both disentangled spectra.
 Tamajo, Pavlovski \& Southworth (2009) applied this idea and found
 light factors derived from genetic forward modelling of H$\gamma$ in the eclipsing
 binary V615\,Per to be in excellent agreement with those found from light curve analysis
 (Southworth et al.\ 2004a). In Fig.\,2 (left panel), renormalised spectra of both
 components of V615\,Per are shown. Generic spectra are fitted
 both for effective temperature and light factor in a constrained mode (sum of light
 factors equals 1). The fits are shown in Fig.\,2 (right panel).
 Light factors derived from light curve analysis and from line-profile fitting of
 disentangled spectra agree within  1.5\%.
 That this method gives reliable results is also evident from the study of
 Pavlovski et al.\ (2009), who made simultaneous constrained fits of the
 three Balmer lines for the primary component of V380\,Cyg.

\end{itemize}

Renormalised disentangled spectra are suitable for spectroscopic analysis by the tools
and methods developed for single stars. Components
of different spectral types and physical characteristics have been analysed. Here is
a taste of some recent studies based on {\sc spd}.
Plaskett's famous star was investigated in detail by Linder et al.\ (2008).
In both disentangled components were disclosed strong abundance anomalies which globally
 corroborate the predictions for evolved, fast-rotating high-mass stars. 
B-type stars are being systematically studied in the series of papers described in
some detail in Sect.\,5. In the field of A-type stars Hareter et al.\ (2008)
 have accomplished a
comprehensive study of the triple system HD\,61199, which photometry from the MOST satellite
revealed to contain a multi-mode $\delta$ Scuti star. In absence of eclipses, the
luminosity contributions of the components were derived by fitting synthetic spectra
to the disentangled spectra assuming a solar composition.
An important contribution in the study of F-type stars in binary systems
has been presented by Clausen et al.\ (2008).

\section{Chemical evolution of high-mass close binary stars}

In the last decade theoretical stellar evolutionary models, particularly
for higher masses, were improved considerably
with the inclusion of rotation and magnetic fields
 These effects
have caused substantial changes in the resulting predictions
(c.f.\ recent reviews by Langer et al.\ 2008 and Meynet et al.\ 2009).
Some of these concern evolutionary changes in the chemical composition of
stellar atmospheres. Due to the CNO cycle in the core of high-mass stars some
elements are enhanced, such as helium and nitrogen, and some are depleted,
like carbon and to a lesser extent oxygen. Rotational mixing is predicted to act so
efficient that changes in the atmospheric composition should be
identifiable whilst the star is still on the main sequence (MS).

Empirical constraints on these processes remain hard to come by.
Analysis of detached eclipsing binaries (dEBs) is vital for specifying
empirical constraints on the properties of high-mass stars, since they are
the primary source of directly measured stellar properties (Andersen 1991;
Torres, Andersen \& Gim\'{e}nez 2009).

In Pavlovski \& Hensberge (2005), it was shown that properly renormalised
disentangled spectra yield abundance measurements comparable in accuracy to
those derived for single stars.  In a series of papers we aim to
calibrate the abundance patterns and chemical evolution of high-mass
stars by analysing dEBs (Pavlovski \& Southworth 2009, Pavlovski et al.\
2009). The core of our analysis is {\em spectral disentangling}, which
reveals the individual spectra of the components. The resulting disentangled
spectra also have a much higher S/N than the original observations,
so are well suited to chemical abundance analysis. Also,
the strong degeneracy between effective temperature and surface
gravity is not a problem for dEBs because surface gravities can be
measured directly and to high accuracy (0.01 dex or better), which is not possible
for single stars.

Binaries studied so far include V578\,Mon (Pavlovski \& Hensberge 2005),
V453\,Cyg (Pavlovski \& Southworth 2009), and V380\,Cyg (Pavlovski et al.\ 2009,
see also Pavlovski et al.\ this Volume).
The components of these binaries have masses in the range 8--15\,M$_\odot$,
and effective temperatures in the range 22--30\,kK, thus are all
early B-type stars. From evolutionary considerations, we find that they are spread throughout MS,
from positions close to ZAMS up to TAMS, like the primary component
of V380\,Cyg. The latter star is an excellent candidate for probing
rotational evolutionary models. Abundance analysis has shown
that all stars in the binaries studied
so far cluster around normal abundances, including the evolved
component in V380~Cyg that has reached already the end of the 
core H-burning phase. This goes against the expectations from rotational
models for the low-end of the high-mass range  (Ekstr\"{o}m et al.\ 2008). 
Both an increase in rotational velocity, and changes in helium and CNO elements,
are predicted by the models.
Recent calculations of the chemical evolution of stars
in close binaries by De Mink et al.\ (2009) have shown that, contrary to
the short period binaries, surface abundance changes should be small for
the relatively long orbital period of V380\,Cyg (12.4\,d). We are currently
engaged in extending the sample of close binaries for which such
comparisons can be made.

\section{Conclusion}

{\sc spd} techniques make possible improvement in deriving emprical data
from binary and multiple systems. Moreover, renormalised reconstructed
spectra can be further analysed and eventually photospheric elemental
abundances can be derived. A new window has opened for probing stellar
evolutionary models by studing the chemical evolution of the components
of binary and multiple stars. To paraphrase the subtitle in Clausen et al.\ (2008)
`{\sl a new standard has been set}' in the study of binary and multiple
stars.

\acknowledgements

We are thankful to Dr.~J.~Southworth for  comments on the first
 version of this paper. Also, we would like to acknowledge collaboration 
with Dr.~S. Iliji\'{c}, and Messrs. E. Tamajo and V. Kolbas at the University 
of Zagreb.


\begin{thebibliography}{}

\bibitem[]{} Andersen, J. 1991, A\&AR, 3, 91
\bibitem[]{} Bagnuolo, W.G., \& Gies, D.R. 1991, ApJ, 376, 266
\bibitem[]{} Bailey, J.A. 1998a, Proc. SPIE, 3355, 932
\bibitem[]{} Bailey, J.A. 1998b, MNRAS, 301, 161.
\bibitem[]{} Baki\c{s}, V., Baki\c{s}, H., Demircan, O., \& Eker, Z. 2008, MNRAS, 385, 381
\bibitem[]{} Baki\c{s}, V., Baki\c{s}, H., Eker, Z., \& Demircan, O. 2007, MNRAS, 382, 609
\bibitem[]{} Baki\c{s}, V., Budding, E., Erdem, A., et al.\ 2006,
          MNRAS, 370, 1935
\bibitem[]{} Boyajian, T.S., Gies, D.R., Helsel, M.E., et al. 2006, ApJ, 646, 1209
\bibitem[]{} Bonanos, A. 2009, ApJ, 691, 407
\bibitem[]{} Clausen, J.V., Torres, G., Bruntt, H., et al. 2008, A\&A, 487, 1095
\bibitem[]{} De Mink, S.E., Cantiello, M., Langer, N., et al. 2009, A\&A, 497, 243
\bibitem[]{} Do\u{g}ru, D., Erdem, A., Do\u{g}ru, S., \& Zola, S. 2009, MNRAS, 397, 1647
\bibitem[]{} Ekstr\"{o}m, S., Meynet, G., Maeder, A., \& Barblan, F. 2008, A\&A, 478, 467
\bibitem[]{} Ferluga, S., Floreano, L., Bravar, U., \& B\'{e}dalo, C. 1997, A\&AS, 121, 201
\bibitem[]{} Fr\'{e}mat, Y., Lampens, P., \& Hensberge, H. 2005, MNRAS, 356, 545
\bibitem[]{} Garcia, P.J.V., Thi\'{e}baut, E., \& Bacon, R. 1999, A\&A, 346, 892
\bibitem[]{} Gies, D.R. 2004, ASP Conf.\ Ser., 318, 61
\bibitem[]{} Gonz\'{a}lez, J.F., \& Levato, H. 2006, A\&A, 448, 283
\bibitem[]{} Gonz\'{a}lez, J.F., Hubrig, S., Nesvacil, N., \& North, P. 2006, A\&A, 449, 327
\bibitem[]{} Griffin, R., \& Griffin, R. 1986, JApA, 7, 45
\bibitem[]{} Griffin, R.E.M., \& Griffin, R.F. 2009, MNRAS, 394, 1393
\bibitem[]{} Groenewegen, M.A.T., Decin, L., Salaris, M., \& De Cat, P. 2007, A\&A, 463, 579
\bibitem[]{} Grundahl, F., Clausen, J.V., Hardis, S., \& Frandsen, S. 2008, A\&A, 492, 171
\bibitem[]{} Grundstrom, E.D., Gies, D.R., Hilling, T.C., et al. 2007, ApJ, 667, 505
\bibitem[]{} Hadrava, P. 1995, A\&AS, 114, 393
\bibitem[]{} Hadrava, P. 1997, A\&AS, 122, 581
\bibitem[]{} Hadrava, P. 2004, Publ. Astron. Inst. Acad, Sci. Czech Rep., 92, 15
\bibitem[]{} Hadrava, P. 2009, A\&A, 494, 399
\bibitem[]{} Hareter M., Kochukhov O., Lehmann H., et al. 2008, A\&A, 492, 185
\bibitem[]{} Harmanec, P., Hadrava, P., Yang, S., et al. 1997, A\&A, 319, 867
\bibitem[]{} Harmanec, P., Mayer, P., Pr\v{s}a, A., et al. 2007, A\&A, 463, 1061 
\bibitem[]{} Harries, T.J., Hilditch, R.W., Howarth, I.D., 2003, MNRAS, 339, 157   
\bibitem[]{} Hensberge, H., \& Pavlovski, K. 2007, IAU Symp., 240, 789
\bibitem[]{} Hensberge, H., Iliji\'{c}, S., \& Torres, K.B.V. 2008, A\&A, 522, 311
\bibitem[]{} Hensberge, H., Pavlovski, K., \& Verschueren, W. 2000, A\&A, 358, 553
\bibitem[]{} Hilditch, R.W. 2007, Observatory,  125, 72
\bibitem[]{} Holmgren, D.E., Hadrava, P., Harmanec, P., et al. 1999, A\&A, 345, 855
\bibitem[]{} Hynes, R.I., Maxted, P.F.L. 1998, A\&A, 331, 167
\bibitem[]{} Iliji\'{c}, S. 2004, ASP Conf. Ser., 318, 107
\bibitem[]{} Iliji\'{c}, S., Hensberge, H.,  \& Pavlovski, K. 2001a,
                             Lecture Notes in Physics, 573,  267
\bibitem[]{} Iliji\'{c}, S., Hensberge, H., \& Pavlovski, K. 2001b, Fizika B, 10, 357
\bibitem[]{} Iliji\'{c}, S., Hensberge, H., Pavlovski, K., Freyhammer, L. 2004,
  ASP Conf. Ser., 318, 111
\bibitem[]{} Langer, N., Cantiello, M., Yoon, S., et al. 2008, IAU Symp., 250, 167
\bibitem[]{} Lee, C.-U., Kim, S.-L., Lee, J.W., et al. 2008, MNRAS, 389, 1630
\bibitem[]{} Linder, N., Rauw, G., Martins, F., et al. 2008,  A\&A, 489, 713
\bibitem[]{} Mahy, L., Naz\'{e}, Y., Rauw, G., et al. 2009, A\&A, 502, 937
\bibitem[]{} Marchenko, S.V., Moffat, A.F.J., \& Eenens, P.R.J. 1998, PASP, 110, 1416
\bibitem[]{} Mayer, P., Harmanec, P., Nesslinger, S., et al. 2008, A\&A, 481, 183
\bibitem[]{} Meibom, S., Grundahl, F., Clausen, J.V., et al. 2009, AJ, 137, 5086
\bibitem[]{} Meynet, G., Chiappini, C., Georgy, C., et al. 2009, IAU Symp. 254, 325
\bibitem[]{} Napiwotzki, R., Chrislieb, N., Drechsel, H., et al. 2003, The Messenger, 112, 543
\bibitem[]{} North, P., Gauderon, R., \& Royer, F. 2009, IAU Symp.\ 256, 57
\bibitem[]{} \"{O}zkarde\c{s}, B., Erdem, A., \& Baki\c{s}, V. 2009, New Astronomy, 14, 461
\bibitem[]{} Pavlovski, K., \& Hensberge, H. 2005, A\&A, 439, 309
\bibitem[]{} Pavlovski, K., \& Southworth, J. 2009, MNRAS, 394, 1519
\bibitem[]{} Pavlovski, K., Tamajo, E., Koubsk\'{y}, P., Southwroth, J., Yang, S., \& Kolbas, V.
     2009, MNRAS, in press ({\sf arXiv:0908.0351})
\bibitem[]{} Penny, L.R., Ouzts, C., \& Gies, D.R. 2008, ApJ, 681, 554
\bibitem[]{} Peters, G.J., Gies, D.R., Grundstrom, E.D., \& McSwain, M.V. 2008, ApJ, 686, 1280
\bibitem[]{} Porter, J.M., Oudmaijer, R.D., \& Baines, D. 2004, A\&A, 428, 327
\bibitem[]{} Rauw, G., Naz\'{e}, Y., Fern\'{a}ndez Laj\'{u}s, E., et al. 2009, {\sf arXiv:0906.2681}
\bibitem[]{} Ribas, I., Jordi, C., Vilardell, F., et al. 2005, ApJ, 635, L37
\bibitem[]{} Simon, K.P., \& Sturm, E. 1994, A\&A, 281, 286
\bibitem[]{} Simon, K.P., Sturm, E., \& Fiedler, A. 1994, 292, 507
\bibitem[]{} Southworth, J., \& Clausen, J.V. 2007, A\&A, 461, 1077                           
\bibitem[]{} Southworth, J., Maxted, P.F.L., \& Smalley, B. 2004a, MNRAS, 349, 547             
\bibitem[]{} Southworth, J., Zucker, S., Maxted, P.F.L., \& Smalley, B. 2004b, MNRAS, 355, 986 
\bibitem[]{} Sturm, E., \& Simon, K.P. 1994, A\&A, 282, 93
\bibitem[]{} Takami, M., Bailey, J., Gledhill, T.M., et al. 2001, 
    MNRAS, 323, 177
\bibitem[]{} Tamajo, E., Pavlovski, K., \& Southworth, J. 2009, A\&A, submitted
\bibitem[]{} Tomkin, J. 1989, Space Science Reviews, 50, 245
\bibitem[]{} Torres, G., Andersen, J., \& Gim\'{e}nez, A. 2009, A\&AR, in press ({\sf arXiv:0908.2624})
\bibitem[]{} Wang, H.J., Wei, J.Y., Shi, J.R., \& Zhao, J.K. 2009. A\&A, 500, 1215
\bibitem[]{} Wheelwright, H.E., Oudmaijer, R.D., \& Schnerr, R.S. 2009, A\&A, 497, 487
\bibitem[]{} Yan, J., Lin, W., \& Hadrava, P. 2008, AJ, 136, 631
\bibitem[]{} Zwahlen, N., North, P., \& Debernardi, Y., et al. 2004, A\&A, 425, L45
\end{thebibliography}
\end{document}